\documentclass[journal=jacsat,manuscript=article]{achemso}

\usepackage{natbib}

\author{Daler R. Dadadzhanov}
\affiliation{ITMO University, 49 Kronverksky Ave., 197101, St. Petersburg, Russia}
\alsoaffiliation{Electrooptics and Photonics Engineering Department, Ben-Gurion University, Beer-Sheva 8410501, Israel}

\alsoaffiliation{Ilse Katz Institute for Nanoscale Science \& Technology, Ben-Gurion University, Beer-Sheva 8410501, Israel}

\email{daler.dadadzhanov@gmail.com}

\author{Tigran A. Vartanyan}
\affiliation{ITMO University, 49 Kronverksky Ave., 197101, St. Petersburg, Russia}

\author{Alina Karabchevsky}
\affiliation{Electrooptics and Photonics Engineering Department, Ben-Gurion University, Beer-Sheva 8410501, Israel}
\alsoaffiliation{Ilse Katz Institute for Nanoscale Science \& Technology, Ben-Gurion University, Beer-Sheva 8410501, Israel}

\email{alinak@bgu.ac.il}

\title {Differential extinction of Vibrational Molecular Overtone Transitions with Gold Nanorods and Non-Trivial Surface Enhanced Near-IR Absorption (SENIRA)}

\begin{document}

%\begin{tocentry}
%
%
%
%\end{tocentry}

\begin{abstract}
 Resonant coupling between plasmonic nanoantennas and molecular vibrational excitations is employed to amplify the weak overtone transitions that reside in the near-infrared. We explore for the first time the differential extinction of forbidden molecular overtone transitions coupled to the localized surface plasmons. We show that a non-trivial interplay between the molecular absorption enhancement and suppression of plasmonic absorption in a coupled system gives rise to orders of magnitude enhancement of the probe molecule differential extinction. Our results pave a road toward a new class of surface enhanced near-infrared absorption-based sensors. 
\end{abstract}

\section{Introduction}

Near-infrared (NIR) spectroscopy focuses on interaction of near-infrared radiation with matter and is an important analytical technique for detection and recognition of chemical substances based on vibrational modes of their molecular constituents in pharmaceutical analysis, food quality determination, non-destructive analysis of biological materials to name a few   \cite{Cen2007TheoryQuality,Manley2014Near-infraredMaterials,Jamrogiewicz2012ApplicationTechnology}. However, molecular overtone bands lying in the NIR spectral region are forbidden in harmonic oscillator approximations \cite{Katiyi2018SiNear-Infrared,karabchevsky2018tuning}. Such bands arise only from the anharmonicity of molecular vibrations which is rather weak \cite{Katiyi2018SiNear-Infrared} leading to the overtone bands with the absorption cross-section of an order of magnitude smaller than the fundamental modes of the same degree of freedom. Here we explore for the first time the mechanism of local field enhancement in molecular overtones. The local field enhancement can be realized with plasmonic materials by means of collective oscillations of free electrons in form of extended surface plasmon-polariton (SPP) in thin metal films \cite{Maier2007Plasmonics:2007.,Klimov2014Nanoplasmonics,karabchevsky2009theoretical, karabchevsky2011fast,karabchevsky2011nanoprecision,karabchevsky2015transmittance} or localized surface plasmon resonance (LSPR) in plasmonic nanoantennas  \cite{Karabchevsky2016TuningNanoparticles,maslovski2018purcell,simovski2015circuit,galutin2017invisibility}.

Enhancement and localization of electromagnetic fields in the close proximity of nanoantennas depend on their material, size, shape and the surrounding media \cite{Karabchevsky2016TuningNanoparticles,Dadadzhanov2018VibrationalNanoantennas}. While exploring the influence of extended surface plasmon on absorption by molecular overtones, we showed that 100 times enhancement can be achieved \cite{Karabchevsky2016StrongPolariton} in guided wave configuration. This enhancement was observed when the absorption band of the molecular vibration N-H was detuned from the plasmonic resonance. In ref. \cite{Shih2016SimultaneousDisks}, authors explored overtones absorption effect with porous gold nanodiscs and ascribed the achieved enhancement to the molecules that occupy hot-spots in the structures. Despite this experimental observation of surface enhanced near-infrared absorption (SENIRA) of molecular overtones with plasmonic nanoantennas, this effect was not explored theoretically.

In this work we theoretically explore yet unclear possibility to enhance absorption by molecular overtone transitions in the near-field of plasmonic nanoantennas such as gold nanorods (GNRs) due to the combination of localized plasmon resonance and lightning rod effect \cite{Li2015plasmon}. 

\section{Theoretical model}

\begin{figure}[h]
\centering
\includegraphics[width=0.5\linewidth]{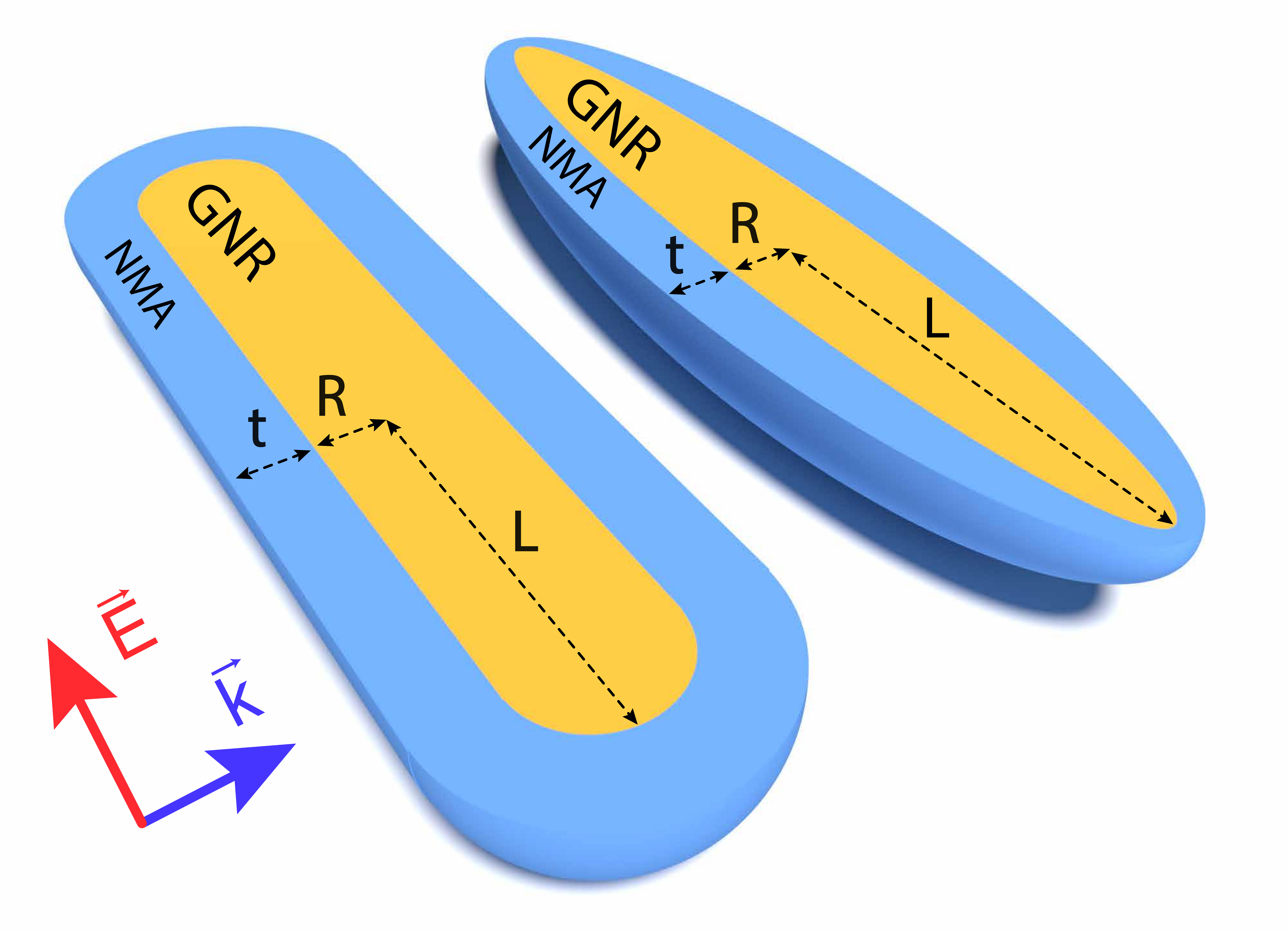}
\caption{Schematics of systems with gold nanorods (left) studied in numerical simulations and nanoellipsoids (right) used in the analytical model. The shells are made of N-Methylaniline (NMA). $L$ and $R$ are semi-major and semi-minor axises of gold nanoparticles respectively, while $t$ is the thickness of molecular shells. The incident wave is polarized along the rod.}
\label{fig:Fig1}
\end{figure}

Fig. \ref{fig:Fig1}  shows the system we study. Weakly absorbing medium, described by the complex permittivity of N-Methylaniline molecule, encapsulates a gold nanorod and nanoellipsoid in a homogeneous shell-like manner. The incident beam is directed perpendicular to the gold nanoparticles as indicated by vector \textit{\textbf{k} }and polarized along the gold nanoparticles.

We study the contribution of GNR parameters in effect of SENIRA by molecular overtones. For this we built an analytical model of a confocal ellipsoidal core-shell nanoparticle. In the framework of quasi-static approximation, we express the absorption, scattering, and extinction cross-sections through the particle polarizability  \cite{Bohren2008AbsorptionParticles,Kelly2003TheEnvironment}:

\begin{equation}
\alpha=\frac{v\left(\varepsilon_{2}-\varepsilon_{m}\right)\left[\varepsilon_{2}+\left(\varepsilon_{1}-\varepsilon_{2}\right)\left(S^{(1)}-f S^{(2)}\right)\right]+f \varepsilon_{2}\left(\varepsilon_{1}-\varepsilon_{2}\right)}{\left(\left[\varepsilon_{2}+\left(\varepsilon_{1}-\varepsilon_{2}\right)\left(S^{(1)}-f S^{(2)}\right)\right]\left[\varepsilon_{m}+\left(\varepsilon_{2}-\varepsilon_{m}\right) S^{(2)}\right]+f S^{(2)} \varepsilon_{2}\left(\varepsilon_{1}-\varepsilon_{2}\right)\right)}
\end{equation}

where \textit{$S^i$} ($i$=1,2) are the geometrical factors of the core and the shell in the direction of the polarization (Fig.\ref{fig:Fig1}); \textit{$\epsilon_1$}, \textit{$\epsilon_{2}$} ,\textit{$\epsilon_m$}  are the frequency-dependent dielectric permittivity function of the gold core, the molecular shell and surrounding media, correspondently; \textit{v} is the full volume of the nanoparticle with the shell and  \textit{f} is the ratio of the inner core volume to $v$.

As the input parameters we consider the core and shell semi-axises and the frequency-dependent dielectric permittivities of the metal core, the shell and the surrounding medium, which was considered to be air. 

The N-Methylaniline (NMA) was chosen as a representative probe-molecule example of an organic molecule that possesses overtone bands in the NIR spectral range \cite{Katiyi2017FigureSpectroscopy,Karabchevsky2016GiantChip,karabchevsky2018tuning}. The absorption bands at wavelengths of 1494 nm and 1676 nm are associated with the first overtones of N-H and C-H stretching modes. These bands are accompanied by the anomalous dispersion regions as it follows from the Kramers-Kronig relations and presented in Figure 7d in Ref. \cite{Katiyi2017FigureSpectroscopy}.

\section{Results and discussion }
First, we analyzed how the LSPR position depends on the analyte shell thickness, $t$. Since the enhanced near-field rapidly decays with the distance from the surface, effective interaction is possible here only at distances comparable to the nanoantenna dimensions. In addition, the aspect ratio of the nanoantenna should provide the resonant interaction between the longitudinal plasmon and an overtone excitation. Therefore, we choose the semi-minor axis of the gold nanoellipsoid as 5 nm, while varying the semi-major axis until the LSPR band overlaps with an overtone mode. 

\begin{figure}[h]
\centering
\includegraphics[width=0.45\linewidth]{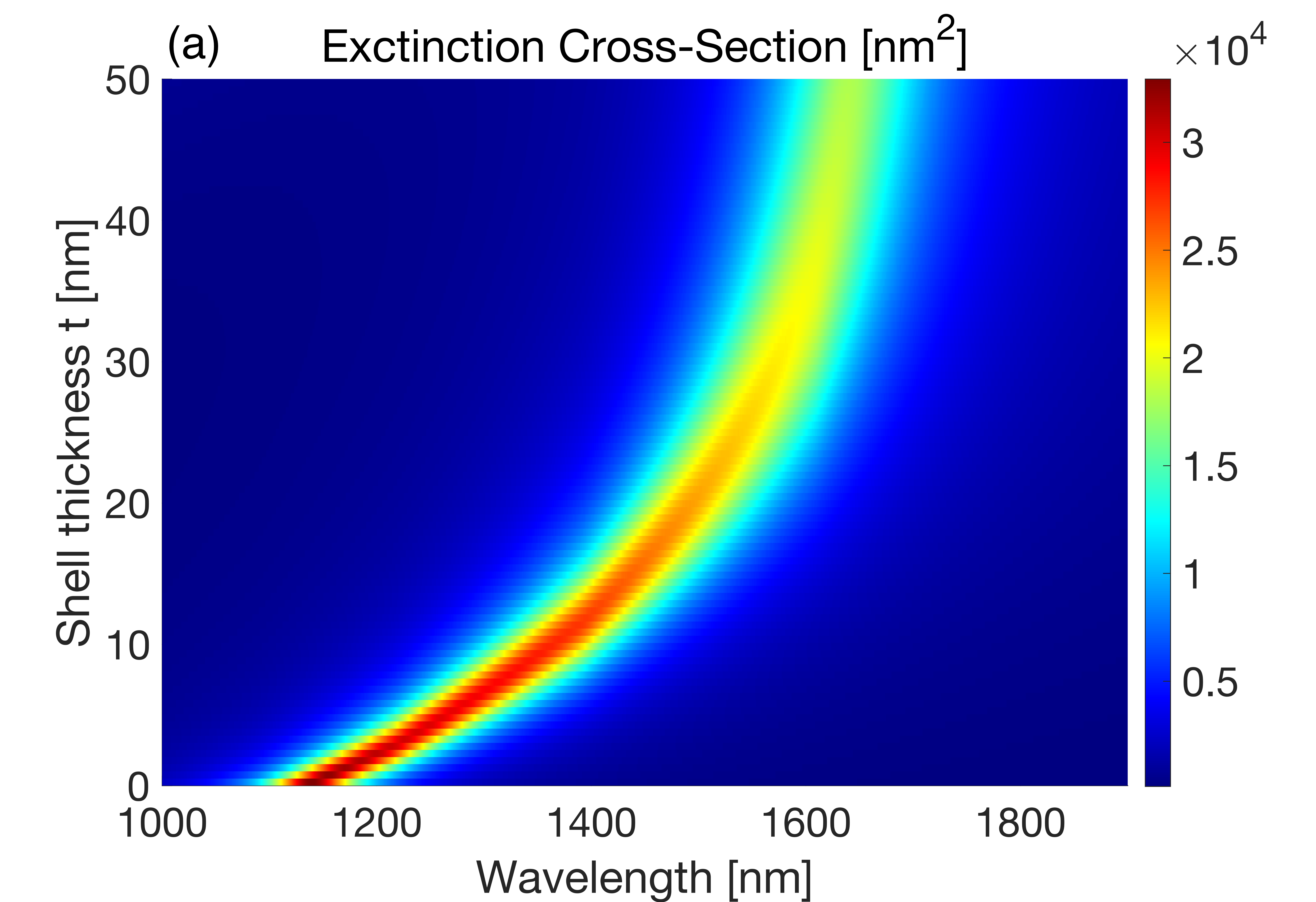}
\includegraphics[width=0.45\linewidth]{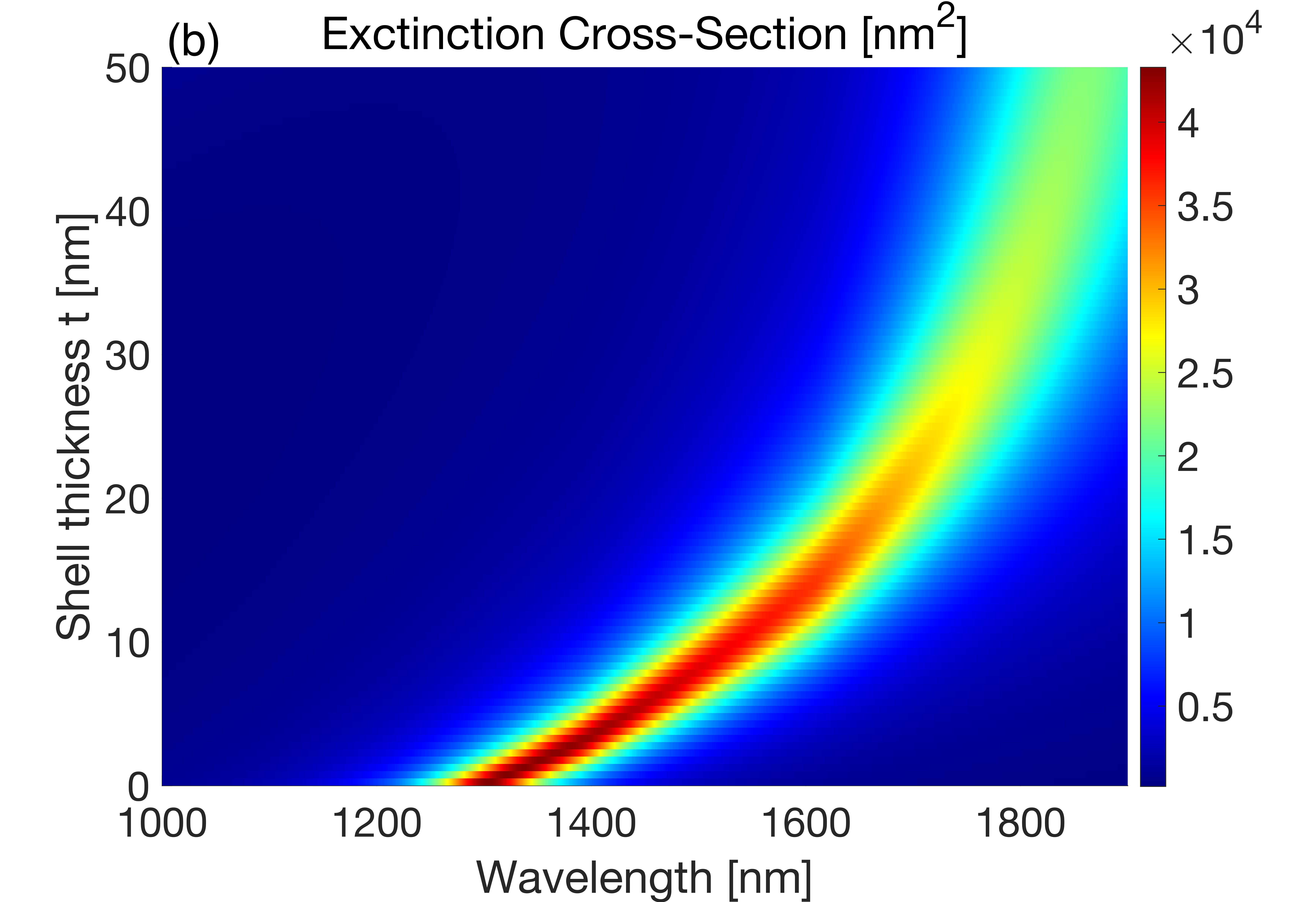}
\caption{Extinction cross-sections of gold nanoellipsoids with NMA shells of different thicknesses. \textbf{(a)} the semi-major axis of the gold core is $L$ = 55.9 nm \textbf{(b)} the semi-major axis of the gold core is $L$ = 68.1 nm. The semi-minor axis is $R$ = 5 nm in both cases. }
\label{fig:Fig2}
\end{figure}

For this, we calculated extinction cross-sections of gold nanoellipsoids covered by thin shells of NMA in the form of confocal ellipsoids. Fig. \ref{fig:Fig2}a shows the extinction cross-section of GNR as a function of the NMA shell thickness. The semi-major axis of the gold core is $L$ = 55.9 nm that leads to the exact resonance with the first overtone of N-H mode when the shell thickness is $t$ = 20 nm. Fig. \ref{fig:Fig2}b  shows the same dependence when the semi-major axis of the gold core is $L$ = 68.1 nm that leads to the exact resonance with the first overtone of C-H mode when the shell thickness is $t$ = 20 nm. The long wavelength shift of plasmon bands as a function of the shell thickness $t$ is rather strong at small shell thicknesses for $t < 40$ nm but saturates at shell thicknesses larger than $ t > 40 $ nm. 

\begin{figure}[h]
\centering
\includegraphics[width=1\linewidth]{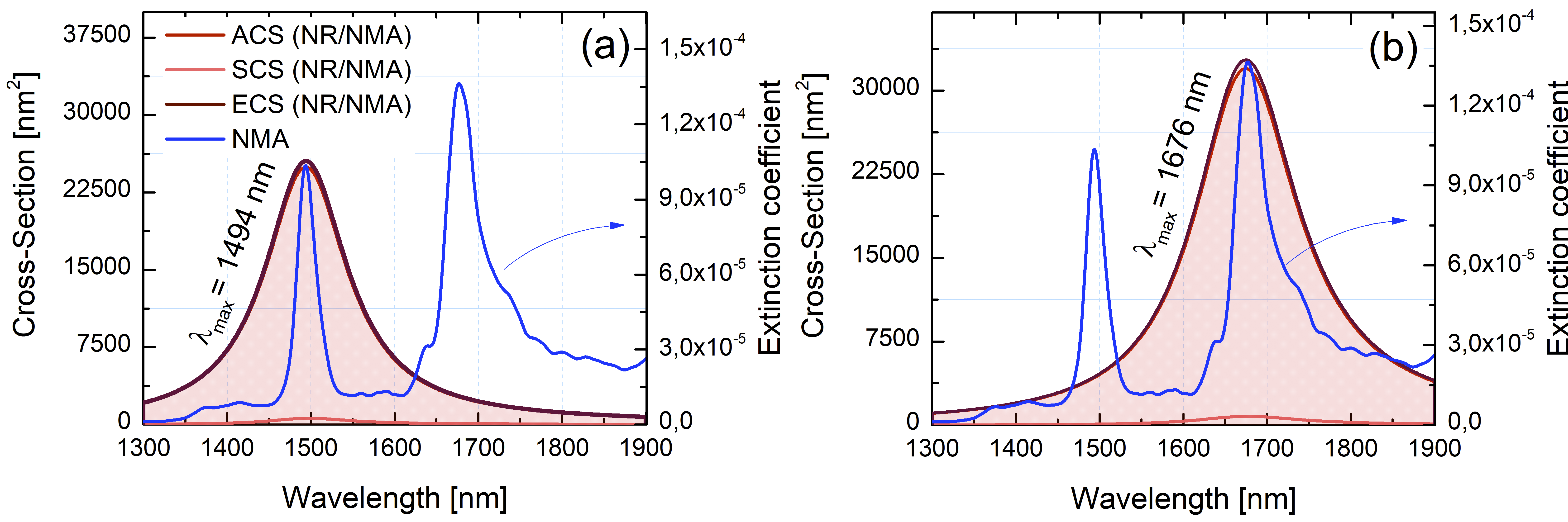}
\caption{Extinction (\textit{brown}), absorption (\textit{red}) and scattering (\textit{pink}) cross-sections of gold nanorods with NMA shell. The nanorod diameter is set to 10 nm for \textbf{(a)} $L$ = 49.9 nm, \textbf{(b)} $L$ = 60.6 nm. The thickness of the NMA molecular shell is homogeneous and equals to $t$ = 20 nm. Extinction coefficient of NMA (\textit{blue}) is also shown for comparison. }
\label{fig:Fig3}
\end{figure}

As proof-of-concept numerical simulations we built numerical model with COMSOL Mul- tiphysics 5.4 software and show the tuning of the plasmon bands of GNR with the NMA overtone bands. Fig. \ref{fig:Fig3} shows calculated extinction (ECS), absorption (ACS) and scattering (SCS) cross-section of gold nanorods with NMA shell for $L = 49.9$ nm (Fig. \ref{fig:Fig3}a) and for $L = 60.6$ nm (Fig. \ref{fig:Fig3}b). The nanorod diameter is 10 nm. We choose the length of GNR such that it overlaps with the overtone bands of N-H located at 1494 nm and C-H located at 1676 nm. Considering the results presented in Fig. \ref{fig:Fig3} one concludes that extinction is governed by absorption, while the scattering contribution is negligible.

The advantage of using GNR becomes evident when the concept of differential extinction is employed  \cite{Karabchevsky2016StrongPolariton}. Experimentally, the differential absorption can be realized by comparing the extinction cross-section of a GNR surrounded by the analyte shell with that of a GNR surrounded by a shell of non-absorbing material that mimic only the mean value of the analyte’s refractive index. Thus, the difference between cross-sections of absorbing and non-absorbing materials represents the influence of the analyte absorption and anomalous dispersion on the LSPR intensity and spectral position. On the other hand, it includes also the influence of the LSPR on the analyte absorption. 

Quantitatively, \textit{differential extinction}, DE, as \cite{Dadadzhanov2018VibrationalNanoantennas,Karabchevsky2016StrongPolariton}:
\begin{equation}
DE=\sigma_{ext}^{NR/NMA}-\sigma_{ext}^{NR/NMA^{*}}
\end{equation}

where the first term $\sigma_{ext}^{NR/NMA}$ represents the extinction cross-section of GNR with NMA shell, while the second term $\sigma_{ext}^{NR/NMA^{*}}$ rrepresents the same value with NMA replaced by a dummy medium of constant dielectric permittivity. Fig.4 shows the calculated DE in spectral ranges of the first overtones of the N-H and C-H stretching modes. Interestingly, the sign of the wavelength dependent DE alternates in both cases. 

\begin{figure}[h]
\centering
\includegraphics[width=0.7\linewidth]{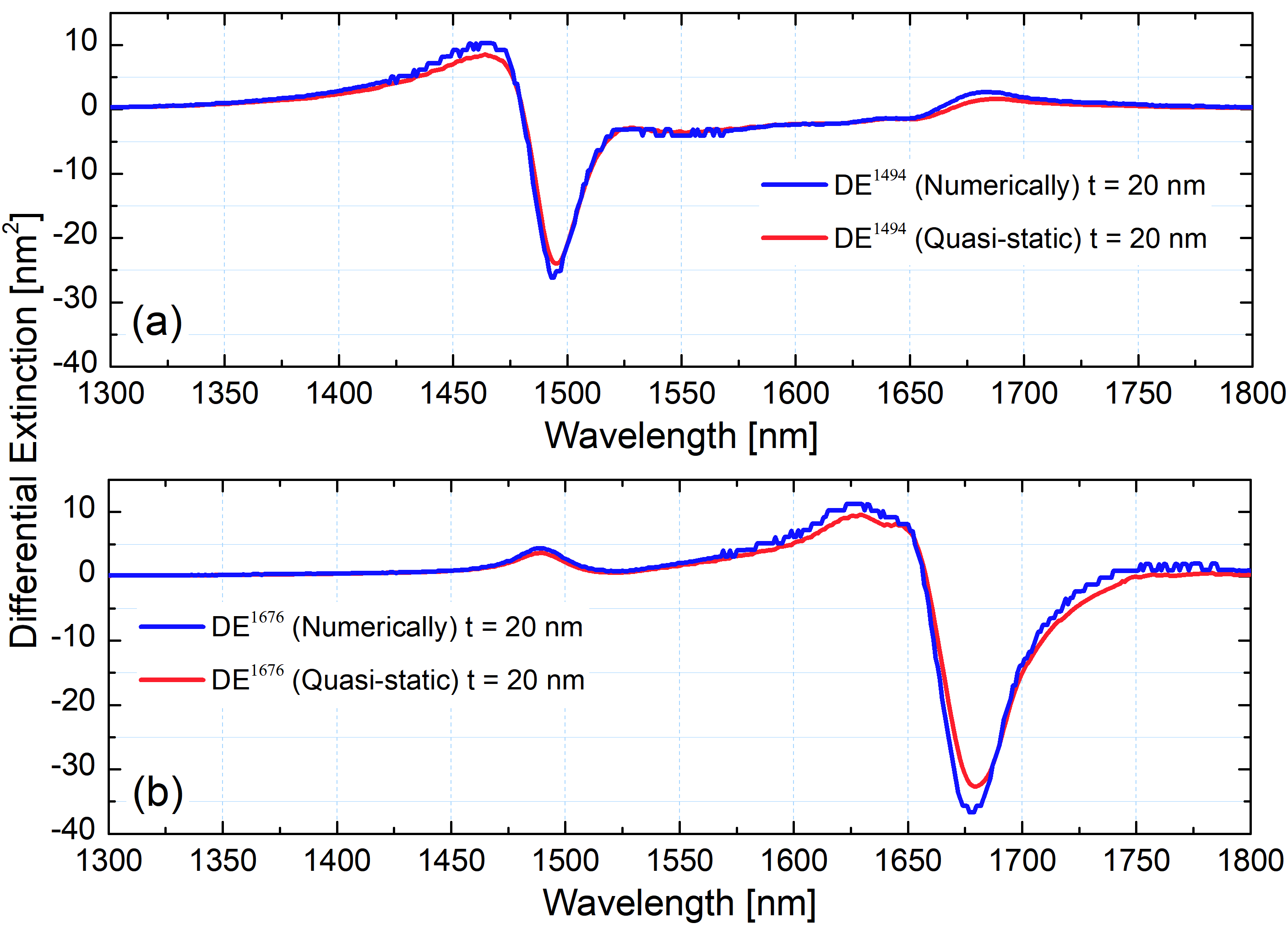}
\caption{Comparative analysis of differential extinction (DE) spectra of gold nanorod and nanoellipsoid with NMA shells: \textbf{(a)} with semi-major axises $L$ = 49.9 (nanorod) and $L$ = 55.9 nm (\textit{nanoellipsoid}); \textbf{(b)} $L$ = 60.6 (\textit{nanorod}) and $L$ = 68.1 nm (\textit{nanoellipsoid}). Blue curves correspond to numerically calculated results (\textit{nanorod}) while the red curves correspond to results obtained in the quasi-static approximation (\textit{nanoellipsoid}) }
\label{fig:Fig4}
\end{figure}

Fig. \ref{fig:Fig4} shows the calculated DE in spectral ranges of the first overtones of the N-H and C-H stretching modes. Interestingly, the sign of wavelength dependent DE alternates in both cases. 

We choose the aspect ratio of nanorods for the Fig.\ref{fig:Fig4} as $L/R$ = 9.98 and the aspect ratio of nanorods for the Fig.4b as $L/R$ = 12.12. Numerically calculated DEs (blue) are very well reproduced by the DEs obtained in the quasi-static approximation (red) (Eq.1) provided the aspect ratios of the GNRs are adjusted to match the plasmon resonance with the corresponding overtone ($L/R$ = 11.18 Fig. \ref{fig:Fig4}a and $L/R$ = 13.62 in Fig. \ref{fig:Fig4}b). It is important to note that in the case exact resonance between the plasmon in the GNR and the molecular overtone transition the sign of DE alternates. Contrary to that in the non-resonant case DE is strictly positive. It may be clearly seen in Fig. \ref{fig:Fig4} for C-H overtone transition at 1676 nm when the plasmon in the nanorod is tuned on 1494 nm (Fig. \ref{fig:Fig4}a) and for N-H overtone transition at 1494 nm when the plasmon in the nanorod is tuned on 1676 nm. 

\begin{figure}[h]
\centering
\includegraphics[width=0.75\linewidth]{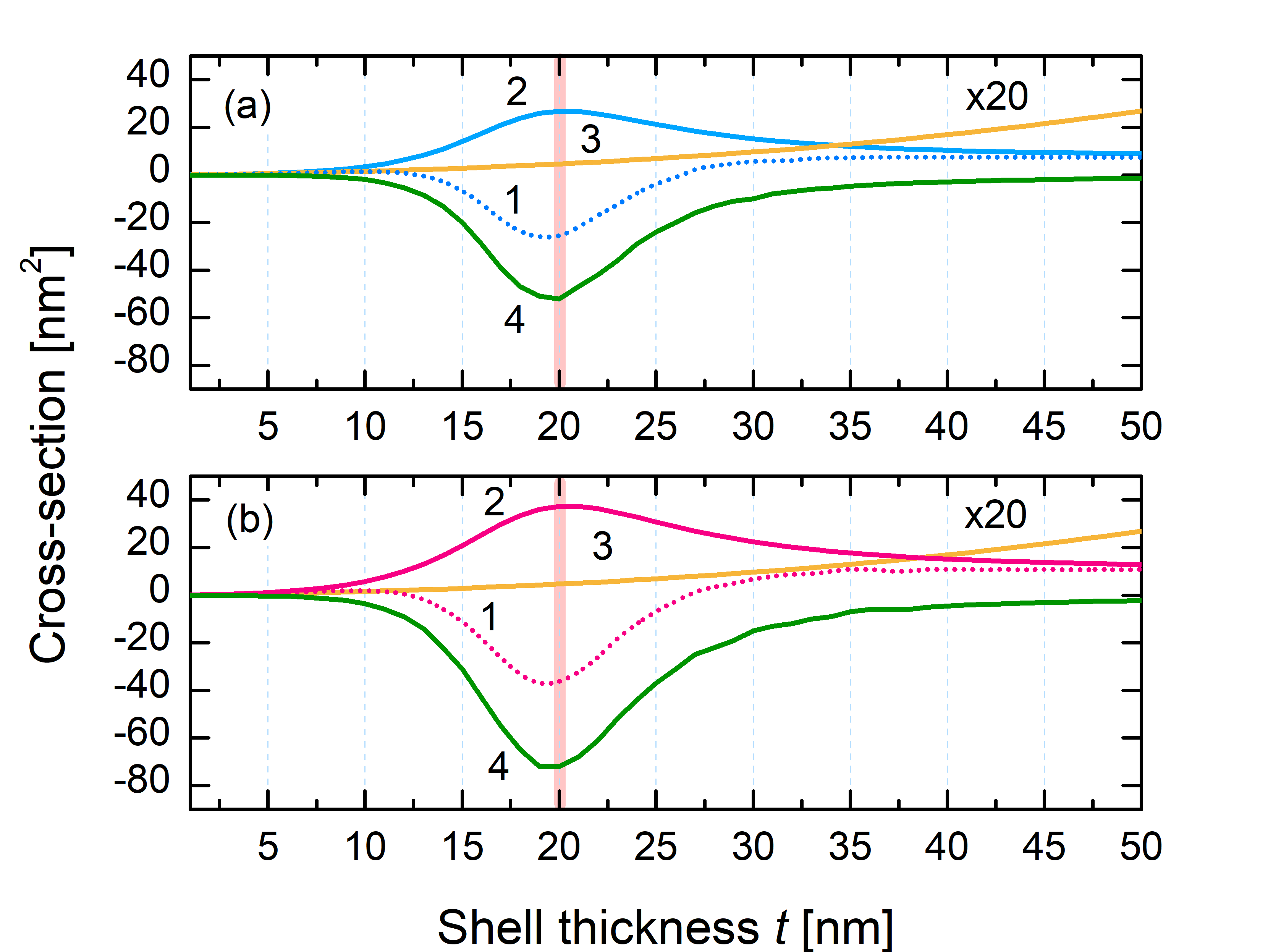}
\caption{\textbf{(a)} Optical cross-section as function of shell thickness of NMA: (1) differential extinction, (2)  absorption cross-section of NMA shell encapsulating GNR, (3) extinction cross-section of the NMA shell without the GNR, and (4) the difference between ACSs of the GNR -with and -without the NMA shell GNR semi-major axis (a)(a)$L$ is 49.9 nm and the wavelength is set to 1494 nm. Absorption cross-section of the NMA shell when the GNR is absent is multiplied by 20. \textbf{(b)} same graphs as in subplot \textbf{(a)} for the case when the wavelength is set to 1676 nm, while $L$ is 60.6 nm.}
\label{fig:Fig5}
\end{figure}

To explore the role of GNRs in the detectivity enhancement of small amounts of NMA, extinction cross-sections of pure NMA shells (without GNR) were compared with the DE. Fig.  \ref{fig:Fig5}  shows the dependence of both values on the NMA shell thickness. When the resonance conditions are met, the DE values exceed the extinction cross-sections of the pure NMA shells by two orders of magnitude. In particular, the first overtone of N-H stretching mode located at 1494 nm is enhanced 114 times, while the first overtone of C-H stretching mode located at 1676 nm is enhanced 135 times. Fig. \ref{fig:Fig5}a shows variations in cross-sections vs. shell thickness for $\lambda=1494$ nm and the GNR semi-major axis is equal 49.9 nm. The resonance conditions for the plasmon excitation are met when the shell thickness is equal to 20 nm. Similarly, the optical properties in the form of ECS and ACS as function of the shell thickness are presented in Fig. \ref{fig:Fig5}b whereby $\lambda=1676$ nm  with GNR semi-major axis $L$ = 60.6 nm.

Enhanced absorption in the NMA shell due to the plasmon near-field is accompanied by reduced absorption in the GNR due to the screening effect \cite{Fofang2008PlexcitonicComplexes}. As a matter of fact, neither enhanced absorption in the shell nor the reduced absorption in the core can be observed in the far-field separately. However, they combine favorably leading to very large DE values.

\begin{figure}[h]
\centering
\includegraphics[width=0.45\linewidth]{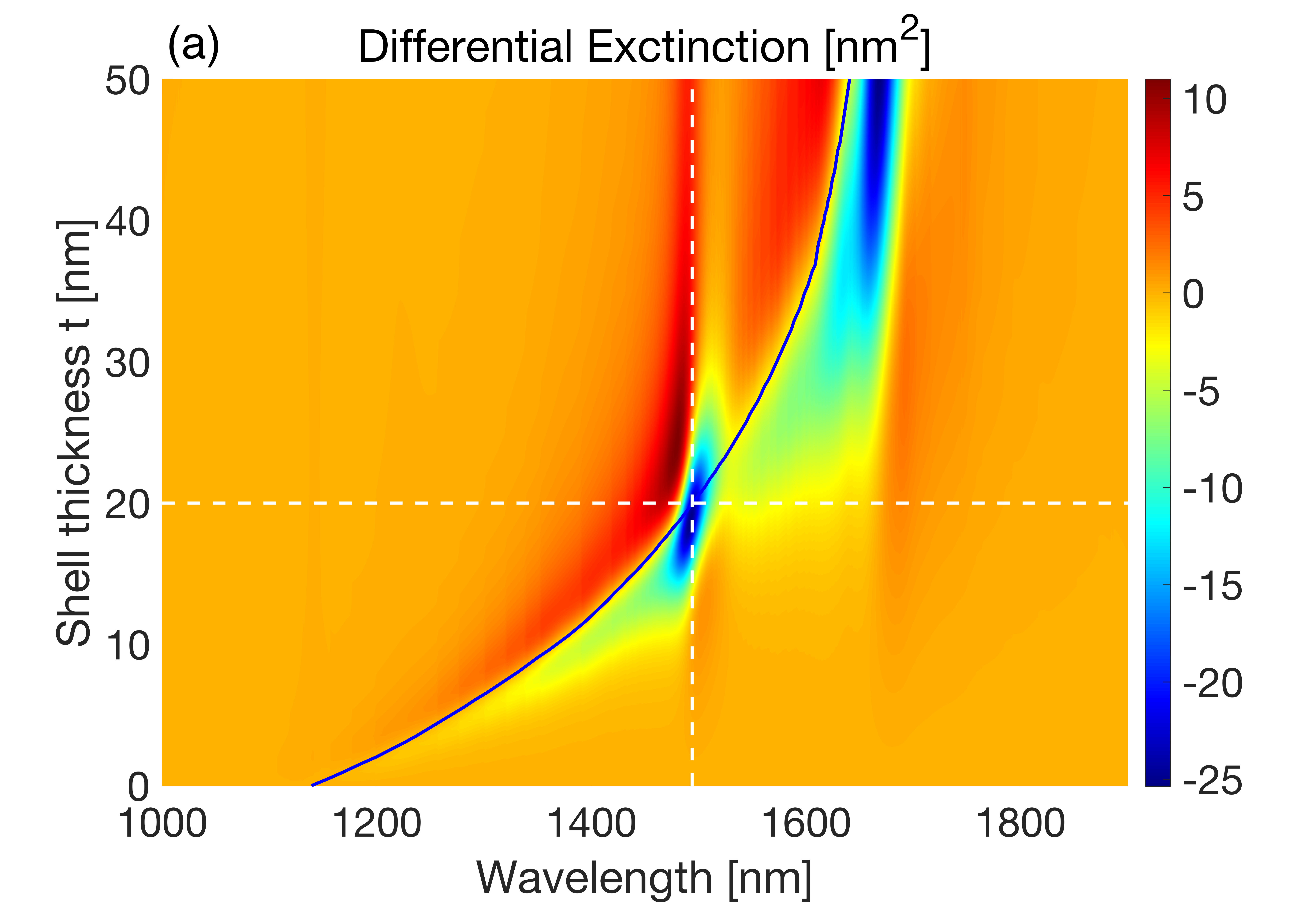}
\includegraphics[width=0.45\linewidth]{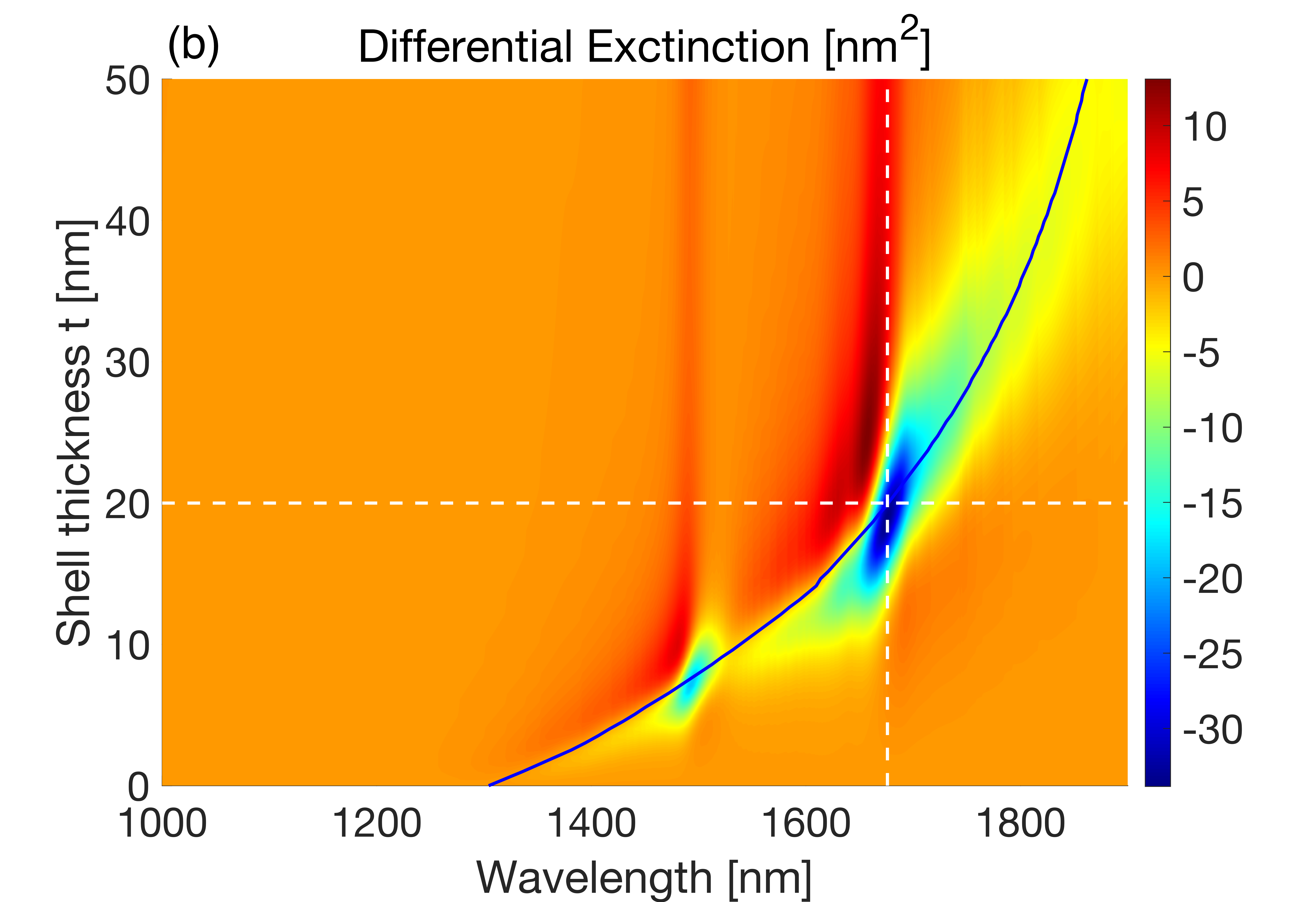}

\caption{\textbf{(a)} Differential extinction (DE) values are given as the functions of the NMA shell thickness and the incident radiation wavelength for GNR with semi-major axis of: $L$ = 55.9 nm \textbf{(a)} and $L$ = 68.1 nm \textbf{(b)}, respectively. The vertical dashed lines show the position of two overtone bands, while the horizontal dashed lines mark the shell thickness ($t$ = 20 nm) that leads to tuned plasmon resonance with the corresponding overtone band. The dark curve is drawn through the maxima of the plasmon resonances.}
\label{fig:Fig6}
\end{figure}

DE dependence on both: the NMA thickness and the incident radiation wavelength based on analytical model is presented in Fig. \ref{fig:Fig6}.  In both plots the dark curve corresponds to the maxima of the LSPR. The vertical dashed lines mark the location of overtone bands, while the horizontal dashed lines correspond to NMA thickness of 20 nm that leads to coincidence of the LSPR in the chosen nanorod with the corresponding overtone band. Inspection of Fig.\ref{fig:Fig6} to the conclusion that the main features of DE already noted in particular cases presented in Fig. \ref{fig:Fig4} and Fig. \ref{fig:Fig5} namely, that the largest absolute value of DE is obtained at the resonance and the sign of this largest DE value is negative are confirmed.

\section{Conclusion}

In conclusion, we explored for the first time the \textit{differential extinction} of forbidden molecular overtone transitions coupled to the localized surface plasmons. We showed that the differential extinction provides  the SENIRA with two orders of magnitude enhancement. The nontrivial consequence of the simulations is that the enhanced absorption in the analyte is accompanied by the reduced absorption in the gold nanorods that overruns the absorption enhancement of the analyte and forms the signal that may be readily sensed in the far-field. Hence, local field enhancement of nanoparticle can result in the considerable sensitivity improvements of overtone spectroscopy in the NIR spectral range. 

\section*{Funding}

This work was supported by the State of Israel-Innovation Authority, Ministry of Economy Grant No. 62045.
The Ministry of Science and Higher Education of Russian Federation (Project 3.4903.2017/6.7).
This work also was financially supported by the Government of Russian Federation, Grant 08-08.
The research described was performed as part of a joint Ph.D. program between the BGU and ITMO University.

\bibliography{ArXiv}

\end{document}